\documentclass[aps,twocolumn]{revtex4}

\usepackage{graphicx}
\usepackage{dcolumn}

\begin{document}
\title{Attosecond resolved charging of clusters}
\author{Ionu\c{t} Georgescu, Ulf Saalmann and Jan M. Rost}
\affiliation{Max Planck Institute for the 
  Physics of Complex Systems\\
  N\"othnitzer Stra{\ss}e 38, 01187 Dresden, Germany}
\date{\today}

\maketitle

\noindent{\sf
  Attosecond laser pulses open the door to resolve microscopic
  electron dynamics in time.   
  Experiments performed include the decay of a core hole
  \cite{drhe+02}, the time-resolved measurement of photo
  ionization \cite{kigo+04} and electron tunneling  
  \cite{ueup+07}.  
  The processes investigated share the coherent character of the
  dynamics involving very few, ideally one active electron.  
  Here, we introduce a scheme to probe dissipative
  multi-electron motion in time.  
  In this context attosecond probing enables one to obtain
  information which is lost at later times and cannot be
  retrieved by conventional methods in the energy domain due to
  the incoherent nature of the dynamics.  
  As a specific example we will discuss the charging of a
  rare-gas cluster during a strong femtosecond pulse with
  attosecond pulses.  
  The example illustrates the proposed use of attosecond pulses
  and suggests an experimental resolution of a controversy about
  the mechanism of energy absorption by rare-gas clusters in
  strong vacuum-ultraviolet (VUV) pulses 
  \cite{wabi+02,sagr03wasa+06,siro04}.
}

Using attosecond pulses for making microscopic time measurements  
involves a unique pump-probe scheme.
In recent experiments \cite{drhe+02,kigo+04,ueup+07}
the system is pumped by an attosecond extreme-ultraviolet (XUV)
pulse and probed by a few-cycle infra-red (IR) pulse. 
In one case \cite{drhe+02,kigo+04}, referred to as attosecond
\emph{streaking}, the final electron momenta
$\vec{p}_\mathrm{final}=\vec{p}_\mathrm{excited} 
+\vec{{\cal A}}(\tau)$, in which the excitation time
$\tau$ is encoded by the instantaneous vector potential
$\vec{{\cal A}}(t)$ of the IR pulse,
is measured. 
In the other case \cite{ueup+07}, the ion yield $P_\mathrm{final}=
\int_{\tau}^\infty\mathrm{d}t'P(t')$ is recorded,
whereby the \emph{nonlinear\/} dependence of the tunnel ionization
$P$ on the instantaneous electric field $\vec{{\cal E}}(t)$ of the IR
pulse allows  one to determine the initial
time $\tau$  with subfemtosecond resolution. 

An alternative imaging method with attosecond resolution which
does not even need ultrashort pulses has been proposed and
demonstrated \cite{itle+04}: 
Using high harmonics generated by illuminating an aligned
molecule with an intense femtosecond laser pulse, the Fourier
transform  of an electronic \emph{amplitude\/} could be imaged
and the corresponding spatial amplitude tomographically
reconstructed.  
Although the measured amplitude looks similar to the highest
occupied molecular orbital of the illuminated molecule this is
certainly not the case since the experiment probes the
multi-electron ground state of N$_2$ which, moreover, is
entangled due to the Pauli principle.  
The details are by now well understood \cite{sago06,pazh+06}. 

Since the new attosecond technique leads into unexplored
territory the first experiments have addressed relatively
simple coherent dynamics: 
one-electron, two-electron, and in the third case a seemingly
simple almost separable case of multi-electron motion. 
One may argue that, at least regarding the pump-probe experiments, the
time-resolved quantities could in principle also be retrieved from an
experiment which exhausts the full parameter space in electron energy.
This is the case since the electron distribution of interest (directly
after the attosecond pulse) remains essentially coherent until it can
be measured (time-resolved or more conventionally, energy-resolved).
Time- and energy-resolved measurements  are 
 to a large extent Fourier related in this context.

Here, we would like to draw attention to another possibly fruitful use
of attosecond pulses in the context of complex systems, exemplified
with rare-gas clusters.  The difference to the simpler electron dynamics
as discussed above lies in the fact that due to dissipation and other
many-particle interaction effects one can by the very nature of the
process not recover time-resolved information from the (energy
resolved) traditional observables.  Hence, time-resolved information
from ultrashort pulses will provide so far unobtainable insight into
the dynamics of those larger, dissipative systems.  Clearly, such
information is dominantly incoherent due to the dissipation processes.

\begin{figure}[t]
  \def\pic#1{\textbf{(#1)}}
  \centering
  \includegraphics[width=0.99\columnwidth]{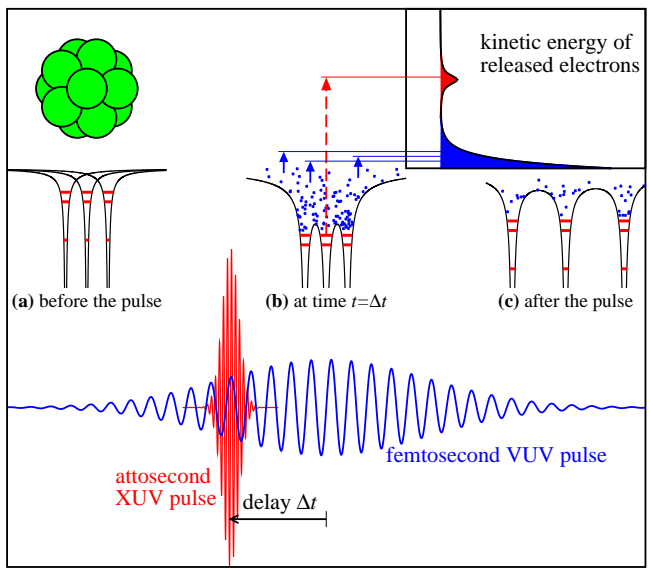}
  \caption{Scheme for the slow-pump (VUV pulse, blue line) fast-probe (XUV
    pulse, red line) time-resolved measurement of the transient charging in a
    multi-ion and multi-electron system.  Before the pulse \pic{a} the electrons
    are localized in atomic orbitals (thick red lines) and the cluster is held
    together by van-der-Waals forces. 
    The initial ionization will lead to a cluster potential
    \pic{b}, allowing for 
    further excitation of bound electrons into the cluster; these quasi-free
    electrons (blue dots) are eventually ionized with low kinetic energies
    (blue arrows).  The attosecond XUV pulse pushes electrons
    from the upper bound states, 
    directly into the continuum by one-photon absorption (dashed red
    arrow).  The measured energy of these electrons traces their instantaneous
    binding energy and thus the transient charge state of the ion and the
    overall charge of the cluster as a function of the time delay $\Delta
    t$ between both pulses.  Finally, after the pulse, the charged cluster
    fragments \pic{c}.}\label{fig:process}
\end{figure}

To be  specific we will propose a pump-probe scenario which is
inverse to the one mentioned above: We expose a small rare-gas cluster
to a femtosecond VUV laser pulse and probe the time-dependent
excitation and ionization dynamics in the cluster, i.\,e., its
charging, by time-delayed attosecond XUV pulses.  
The latter simply kick out electrons, still bound to an ion in
the cluster at that time, and provide in this way information on
the (transient) charging status of the cluster ions, as will be
detailed below and is schematically shown in
Fig.\,\ref{fig:process}. 

We will use for the pump pulse a 100\,fs pulse with a frequency
of 20\,eV similar to the one applied at FLASH in Hamburg in 2002
\cite{wabi+02}.  
The interpretation of the experimental results is still somewhat
controversial 
 \cite{sagr03wasa+06,siro04,jura+05,gesa+07} 
and a future attosecond resolved observation of the charging of
the cluster could discriminate between the different theoretical
scenarios which predict quite different degrees of transient
charging of the cluster. 

\begin{figure}[b]
  \centering
  \includegraphics[width=0.95\columnwidth]{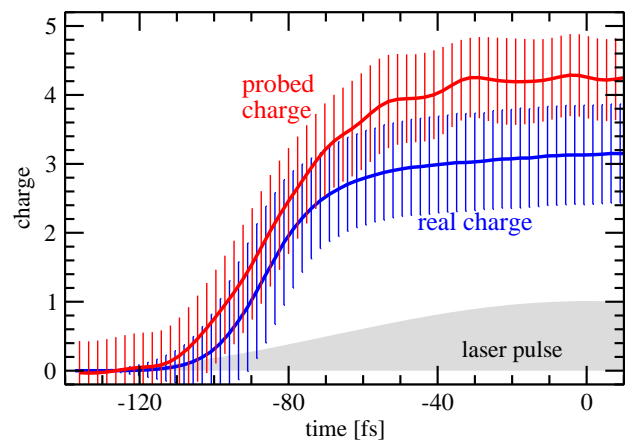}
  \caption{Charging of the Ar$_{13}$ cluster as a function of
    time for a VUV pulse (gray-shaded area) with a photon
    frequency of 20\,eV, an intensity of
    $7\times10^{13}$\,W/cm$^{2}$, a pulse length of 100\,fs
    and the peak at $t=0$.
    The final ionic charge (thick blue line: average, thin blue
    lines: standard deviation) is reached early in the pulse
    during the time intervall from $-100$\,fs to $-60$\,fs.
    Attosecond XUV probing nicely reproduces the charging (red
    lines) by converting the kinetic-energy spectra (Fig.\,3)
    for the  respective time delay $\Delta t$.}\label{fig:charging}
\end{figure}
In our specific example we want to trace the ionization stage of the
cluster ions through the kinetic energy spectra of the photo electrons
from the attosecond pulse.  Three possible problems come immediately
to mind.  They all have to do with the unique relation of the time of
flight spectra of the electrons to the time dependent charging of the
target:
\begin{itemize}
\item[(i)] The attosecond pulse will act perturbatively on the
electrons bound to the ions, but how does one make sure that
the least bound electron is kicked out and therefore identifies
uniquely the present charge state of the ion?  
\item[(ii)] How can one avoid  that the atto-pulse photo
electron  looses its energy
characteristic of the bound state it came from
through inelastic collisions while leaving the cluster?
\item[(iii)] How does
one distinguish between ``normal'' ionized electrons and the atto-pulse
photo electrons?
\end{itemize}
In order to answer these questions and to see if the idea of
slow-pump fast-probe is feasible one has to do a realistic
calculation.  
We base our calculation on previous experience in modeling
cluster dynamics \cite{siro02,saro02,saro03,siro04,gesa+07}.  
As a new element
we add the atto-pulse probe which acts perturbatively (via photo
ionization rates) onto individual ions in their respective charge
state. 
The photo-ionization rates for the argon atom in the XUV region are
easily accessible \cite{co81} since 3s and 3p electrons from the outer
shell can be ionized by single-photon absorption.  Electrons in other
states are not affected by the attosecond XUV pulse.  Quasi-free
electrons, on one hand, have due to their delocalization in the
cluster volume much lower cross sections for ionization.  Core-shell
electrons (1s, 2s, 2p), on the other hand, would require multi-photon
processes which are negligible at the XUV intensity applied.

Altogether, the
well defined coupling of the $n=3$ shell electrons to the attosecond
pulse solves problem (i), although in the case of Argon, indeed, 3s 
and  3p electrons are ionized, as we will see.
Problem (ii) is solved by minimizing the probability of collisions for 
the photo electron using a small Ar$_{13}$ cluster  and a 
high frequency for the atto pulse (fast photo electron).
The latter also solves problem (iii) since the quasi-free electrons form
a well characterized electron plasma whose Maxwellian yields electrons
in the continuum but with energies of only a few eV \cite{laru+05},
which should be very small compared to the energies of the photo
electrons. 
One may think that a slower probe than with attosecond pulses may be 
sufficient for the
charging dynamics.  However, the simulation
(without atto-pulse) suggests that the interesting charging happens
within some 10\,fs, see Fig.\,\ref{fig:charging}.  
Hence, probe pulses of at least subfemtosecond length are
convenient.  
Moreover, the probe pulse must have considerably higher photon
energy to generate photo-electrons, clearly separated in energy
from the cluster electrons, see problem (iii). 
We have used 500\,as pulses with 150\,eV photon energy and a
peak intensity of $1.4\times 10^{15}$\,W/cm$^{2}$.

\begin{figure}[t]
  \centering
  \includegraphics[width=0.99\columnwidth]{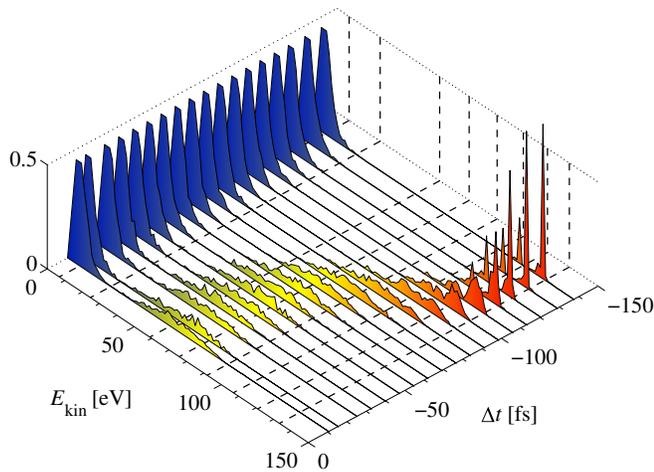}
  \caption{Series of kinetic energy spectra of the released
    electrons for several time delays $\Delta t$ between XUV
    pump and VUV probe pulse. 
    The dashed lines indicate the excess energy $\hbar\omega_\mathrm{xuv} -
    E_\mathrm{ip}(\mbox{Ar}^{q+})$ over the ionization potential
    $E_\mathrm{ip}$ of an isolated $q$-fold charged argon.
    They allow for the identification of the transient charge $q$
    of the mother ion. 
    The low-energy part of the spectrum is independent of
    $\Delta t$.  It is formed by
    quasi-free electrons,  evaporated from the electron plasma 
    which is typically generated by the VUV
    excitation \cite{laru+05}. 
    At a delay $\Delta t = -137$\,fs the XUV
    pulse probes the neutral cluster, as the VUV pump is still
    off.}\label{fig:spectra}
\end{figure}

Figure \ref{fig:spectra} presents a series of electron spectra
with different delays $\Delta t$ between the femto- and the
atto-pulse.   
One can clearly follow the charging of the cluster with
increasing delay of the atto pulse.  
For a time delay $\Delta t=-137$\,fs of the atto pulse, the atomic 
ionization spectrum is obtained (uppermost curve) with the main
peak from the 3p orbitals and a smaller peak from the 3s orbitals.  
Note that the peaks are slightly redshifted with respect to
the expected line from an isolated atom (dashed lines in
Fig.\,\ref{fig:spectra}).   
This is due to the positive background charge of the cluster
against which the electrons must find their way to the
continuum.  
Consequently, the red shift becomes larger for longer delay,
when the cluster charge increases. 
However, apart from the growing red shift and a slight
broadening, the shape of the spectrum remains the same until
a delay of $\Delta t=-105$\,fs: 
There, for the first time a signal at higher charges (lower
kinetic energy) becomes visible which intensifies for the longer
delays shown ($-98$\,fs to $-75$\,fs).  
One notes that after $\Delta t=-70$\,fs the attopulse induced photo electron
distribution remains roughly the same in shape but moves still
to lower kinetic energies for longer time delay.  
This indicates that the internal charging of the cluster has
come to an end, but hot plasma electrons slowly evaporate,
leaving a higher and higher background charge which makes it
more difficult for the photo electrons to escape.  
Note, that the low-energy part of the electrons from the plasma
(blue wings in Fig.\,\ref{fig:spectra}) is not influenced by the 
XUV pulse.

Since for the high-energy part the kinetic energy is given by
$E_\mathrm{kin}=\hbar\omega_\mathrm{xuv}-E_\mathrm{ip}(\text{Ar}^{q+})$,
with the latter the ionization potential of a $q$-fold charged
argon ion, one can measure the average charge from the centre of
mass of the obtained energy distribution.
The measured charge (red in Fig.\,\ref{fig:charging}) leads
the real one (blue).
Within the standard deviation the measured charge is
systematically larger due to the cluster's overall
charge and the excited states of the ions.

We have proposed a scheme to probe the transient charging of a
rare-gas cluster during exposure to a strong VUV pulse with
attosecond pulses. 
This specific example conveys the general idea for a so far not
suggested use of attosecond pulses: 
Initiate a relatively slow excitation process in an extended
system through a femtosecond laser pulse, and probe the
non-stationary, most likely dissipative relaxation dynamics by
time delayed attosecond pulses.

\subsection*{Methods}\noindent
A hybrid quantum-classical approach is used to simulate the
time-dependent dynamics of the cluster ions and electrons under
external light pulses \cite{gesa+07}.  Bound electrons are not
explicitly treated, but can be ionized with probabilities according to
their quantum ionization rates.  Special attention has been paid to
adopt the photo ionization rates to the cluster environment containing
electrons which screen the ion and neighboring ions which lower the
threshold for ionization into the cluster.  Once an electron is
ionized into the cluster, it becomes ``quasi-free'' and is
propagated together with all other charged particles (ions and
electrons) classically under the full 
Coulomb interaction and the dipole coupled electric field 
$\vec{{\cal E}}(t)=\vec{e}_z\sqrt{I_\mathrm{vuv}/I_0}
\sin^2\left(\frac{\pi t}{2.75T_\mathrm{vuv}}\right)
\sin\left(\omega_\mathrm{vuv}t\right)$ of the
VUV pulse which has a peak intensity of $I_\mathrm{vuv}=7\times
10^{13}$\,W/cm$^{2}$ at $\hbar\omega_\mathrm{vuv}=20$\,eV photon
energy and $T_\mathrm{vuv}=100$\,fs pulse length.
The effect of the attosecond pulse is treated in perturbation
theory.  
The intensity $I_\mathrm{xuv} = 1.4\times 10^{15}$W/cm$^{2}$ of
the atto pulse (photon energy $\hbar\omega_\mathrm{xuv}=90$\,eV,
duration $T_\mathrm{xuv}=500$\,as) is chosen such that on
average 0.5 electrons/cluster are photo ionized by the atto
pulse to ensure that the atto electrons do not perturb the
charging situation of the cluster themselves.

\end{document}